\documentclass{elsart}

\usepackage{amssymb}

\def\qb{\bar{q}}
\def\go{\tilde{g}}
\def\sfa{\tilde{f}}
\def\sfb{\bar{\tilde{f}}}
\def\sq{\tilde{q}}
\def\sqb{\bar{\tilde{q}}}
\def\sqlr{\tilde{q}_{\rm L,R}}
\def\stez{\tilde{t}_{1,2}}
\def\sbez{\tilde{b}_{1,2}}
\def\po{\tilde{\gamma}}
\def\zo{\tilde{Z}}
\def\nhoe{\tilde{H}^0_1}
\def\nhoz{\tilde{H}^0_2}
\def\noev{\tilde{\chi}^0_{1-4}}
\def\noi{\tilde{\chi}^0_i}
\def\noe{\tilde{\chi}^0_1}
\def\noz{\tilde{\chi}^0_2}
\def\wo{\tilde{W}^{\pm}}
\def\hoee{\tilde{H}_1^1}
\def\hoez{\tilde{H}_1^2}
\def\hoze{\tilde{H}_2^1}
\def\hozz{\tilde{H}_2^2}
\def\choe{\tilde{H}^{\pm}_1}
\def\choz{\tilde{H}^{\pm}_2}
\def\coez{\tilde{\chi}^{\pm}_{1,2}}
\def\coi{\tilde{\chi}^{\pm}_i}
\def\coe{\tilde{\chi}^{\pm}_1}
\def\coz{\tilde{\chi}^{\pm}_2}
\def\copi{\tilde{\chi}^+_i}
\def\comi{\tilde{\chi}^-_i}
\def\sllr{\tilde{l}_{\rm L,R}}
\def\snl{\tilde{\nu}_{\rm L}}

\def\ms{m_{\tilde q}}
\def\mf{m_{\tilde f}}
\def\mg{m_{\tilde g}}
\def\ghat{\hat{g}_s}
\def\eps{\varepsilon}

 \def\lp{\left. }
 \def\rp{\right. }
 \def\lr{\left( }
 \def\rr{\right) }
 \def\le{\left[ }
 \def\re{\right] }
 \def\lg{\left\{ }
 \def\rg{\right\} }
 \def\lb{\left| }
 \def\rb{\right| }

 \def\met{\rlap{\,/}E_T}

 \def\beq{\begin{equation}}
 \def\eeq{\end{equation}}
 \def\bea{\begin{eqnarray}}
 \def\eea{\end{eqnarray}}

\begin{document}

\begin{frontmatter}



\title{Top quark threshold 
production in $\gamma\gamma$ collisions:
current theoretical issues\thanksref{talk}}
\thanks[talk]{Talk presented at the 
International Workshop on High Energy Photon Colliders
DESY, Hamburg, Germany, June 14-17, 2000.}


\author{A.~A. Penin \thanksref{address}}

\address{II.\ Institut f\"ur Theoretische Physik, Universit\"at Hamburg, \\
Luruper Chaussee 149, D-22761 Hamburg, Germany}

\thanks[address]{Permanent address: 
Institute for Nuclear Research,  60th October Anniversary Pr., 
7a, Moscow 117312, Russia.}

\begin{abstract}
The top quark-antiquark pair threshold
production at future High Energy Photon Colliders is considered 
in view of the  recent  advances in the theoretical  
description  of the nonrelativistic heavy  quark  dynamics.
 
\end{abstract}

\end{frontmatter}



The future   Photon Colliders provide an opportunity 
of experimental study of high energy $\gamma\gamma$ interactions
which can be used for the top quark-antiquark 
pair production. 
Using the laser backscattering method one can obtain 
$\gamma\gamma$ colliding beams with the energy and luminosity 
comparable to those in $e^+e^-$ annihilation \cite{Gin}. 
Theoretical study of the top quark-antiquark 
pair production near the two-particle threshold  \cite{Fad}
is based on the key observation that the relatively large 
electroweak  top quark width $\Gamma_t$ 
and characteristic scale of the 
nonrelativistic Coulomb dynamics 
$\alpha_s^2m_t$ are considerably larger than $\Lambda_{\rm QCD}$ 
and serve  as an effective infrared cutoff for long distance 
nonperturbative strong interaction effects.
This makes perturbative 
QCD applicable for the theoretical description of the threshold
top quark physics. At the same time  numerically
$\Gamma_t\sim \alpha_s^2m_t$ and 
the Coulomb effects are not completely 
dumped by the non-zero top quark width 
that should be properly taken into account.

High quality experimental data that can be obtained in the 
experiments along with a very accurate 
theoretical description of them
make the processes of top-antitop
pair threshold production a promising place
for investigating quark-gluon
interactions. This investigation concerns both general features
of interaction and precise quantitative properties such as
the determination of numerical values of
the strong coupling constant $\alpha_s$,
the top quark mass, and the top quark width.
The main features  of $\gamma\gamma\to t\bar t$ 
threshold production  are rather similar to 
the properties of the $e^+e^-\to t\bar t$ process.
However, the strong interaction  and relativistic
corrections are different for them.
Therefore a simultaneous analysis of
these two processes extends possibilities of studying
fine details of the top quark threshold dynamics.
Moreover, the $S$ and $P$
partial waves of the final state top quark-antiquark pair
produced in $\gamma\gamma$ collisions can be separated
by choosing the same or opposite helicities of the
colliding photons. This gives an opportunity of direct
measurement of the $P$ wave amplitude which is strongly
suppressed in the threshold region in comparison to the $S$ wave one
and provides us with an  additional
independent probe of  the  top quark interactions.

Recently an essential progress has been achieved
in study of the top quark-antiquark 
pair production in $e^+e^-$ annihilation 
reviewed in \cite{gang}. This analysis was
extended to $\gamma\gamma$ collisions in 
the next-to-leading order (NLO)  \cite{PenPiv1}
and next-to-next-to-leading order (NNLO) \cite{PenPiv2}
of the perturbative and relativistic expansion. 
Below we outline  the approach used in these
papers and summarize the obtained results. 
We consider the normalized total cross section
\begin{equation}
R(s)={\sigma(\gamma\gamma\rightarrow t\bar t)\over
\sigma(e^+e^-\rightarrow\mu^+\mu^-)} \, ,
\label{Rgg}
\end{equation}
which can be decomposed into the sum $(R^{++}+R^{+-})/2$ 
of the cross sections $R^{++}$ and $R^{+-}$ for
the colliding photons of the same and opposite helicity
respectively.
Near the threshold 
the heavy quarks are nonrelativistic so that one may consider
the quark velocity $v$ as a small parameter.
An expansion in $v$ may be performed directly in the Lagrangian of QCD by
using the effective theory framework of nonrelativistic QCD (NRQCD)
\cite{CasLep}.
In the effective theory
the dynamics of the heavy quarks is governed 
by the  nonrelativistic 
Schr\"odinger equation and by their multipole interaction to the 
dynamical ultrasoft gluons.
The corrections from harder scales are contained in the Wilson coefficients
leading to an expansion the $\alpha_s$ as well as in the higher-dimensional
operators  corresponding to the expansion in $v$.
In the  threshold region
the cross sections are determined by the imaginary part
of the correlators of the nonrelativistic vector/axial  
quark currents which can be related to the nonrelativistic
Green function $G({\bf x},{\bf y},E)$ and its derivatives at the origin
\[
R^{++}(s)
={24\pi q_t^4N_c \over m_t^2}
\left(C^{++}(\alpha_s)-{1\over 3}{E\over m_t}+\ldots\right)
{\rm Im}G(0,0,E) \, ,
\]
\begin{equation}
R^{+-}(s)={32\pi q_t^4N_c\over m_t^4}(C^{+-}(\alpha_s)
+\ldots)\partial^2_{\bf x\bf y}
{\rm Im}G({\bf x},{\bf y},E)|_{x,y=0}\, ,
\label{Rpm}
\end{equation}
where $N_c=3$, $q_t=2/3$, 
$E=\sqrt{s}-2m_t$ is the energy of a quark pair
counted from the threshold and ellipsis stand
for the high order relativistic corrections.
A symbolic notation  $\partial^2_{\bf x\bf y}$ 
is used for the operator  
that singles out the $P$ partial wave of  $G({\bf x},{\bf y},E)$.
Note that the cross section  $R^{+-}$ is suppressed  by
the factor $v^2$ in comparison to $R^{++}$.
Therefore only  the NLO
corrections to $R^{+-}$ cross section 
is of practical interest.
The nonrelativistic Green function 
satisfies the Schr\"odinger equation
\begin{equation}
\left({H}_C+\Delta{H}-E\right)G({\bf x},{\bf y},E)
=\delta^{(3)}({\bf x}-{\bf y})\, ,
\label{Sch}
\end{equation}
where $H_C$ is the Coulomb Hamiltonian and $\Delta{H}$
stands for the high order corrections in $\alpha_s$ and $v$.
The leading order approximation for the
nonrelativistic Green function 
is obtained with the Coulomb Hamiltonian
and sums up   the singular $(\alpha_s/v)^n$
Coulomb terms  to all orders. The higher order 
terms in the effective  Hamiltonian are known 
up to NNLO including $O(\alpha_s^2)$ perturbative
corrections \cite{Pet}. Corresponding corrections 
to the Coulomb Green function have been obtained in 
\cite{KPP,PinYnd,HoaTeu1,MelYel} and to its derivative 
in \cite{PenPiv1}. The ultrasoft gluons do not contribute
in NNLO.
The perturbative Wilson coefficients   $C^{++}$ and $C^{+-}$
in Eq.~(\ref{Rpm}) are known in NLO.
To complete the NNLO analysis of the $R^{++}$ cross section
the  NNLO
contribution to  the coefficient $C^{++}$ has to be computed. 
Now only the $O(\alpha_s^2)$ 
anomalous dimension \cite{PenPiv2} and the Abelian part \cite{CMY}
of this coefficient   are available.
Because the threshold dynamics is nonrelativistic
and is rather insensitive to
the hard momentum details of top quark
decays the instability of the top quark can be
implemented simply by the complex energy shift 
$E\to E+i\Gamma_t$ in Eq.~(\ref{Sch}). 
This  accounts for the leading imaginary 
electroweak contribution to the leading order NRQCD Lagrangian  
and the most essential features of the physical situation
are caught within this approximation.
However, in the case of $P$ wave production 
the above prescription  is not sufficient
for a proper description of the entire effect of the
non-zero top quark width  and more thorough analysis
is necessary (see \cite{PenPiv1} for detailed discussion).

To summarize,  
the complete NLO analytical  expressions for the 
$R^{++}$ and $R^{+-}$ cross sections
are known and a  bulk  of NNLO corrections to the  
total cross section of  unpolarized or 
equally polarized photons is available.
The cross sections  $R$ in NNLO and  $R^{+-}$ 
in NLO are plotted on Fig.~1. and Fig.~2.
respectively as the functions of energy
for several values of the 
``soft'' normalization point $\mu$ of the
strong coupling constant of the nonrelativistic 
Coulomb problem.
In  the numerical analysis we  neglect
the unknown  non-logarithmic $O(\alpha_s^2)$ 
part of the Wilson coefficient  $C^{++}$.
Taking into account  the result for  the similar coefficient
in the analysis of the photon mediated $t\bar t$  production in  
$e^+e^-$ annihilation obtained in \cite{CM}
we  suppose  this contribution to lead only to a few percent change
of  the overall normalization of the cross section.
The main conclusions we draw from the resuts
of the numerical analysis are: 
(i) the ground state Coulomb  resonance  is 
distinguishable in the cross section $R$
while in  $R^{+-}$ the resonances are completely
smeared out by the top quark width;
(ii) the total cross section of  unpolarized or
equally polarized photons  suffers from the 
large  corrections and is quite sensitive
to the normalization point of the strong coupling constant
in NNLO. 

The importance of the high order corrections
to the cross section has been proved by explicit calculation
of certain classes of the next-to-next-to-next-to-leading
corrections (see \cite{Pen} as a review). In \cite{KniPen1} the retardation
effects caused by the ultrasoft gluons were analysed. 
The leading logarithmically
enhanced $O(\alpha^3_s)$ corrections  
to the resonance energy and normalization 
have been computed in \cite{KniPen2}.
The Abelian part of the subleading logarithmic corrections  
was obtained in \cite{KniPen3} in the context of positronium lifetime
calculation. On the basis of this analysis and from
the normalization scale dependence of the NNLO result
we estimate the high order
corrections to the cross section to be at least
$10\%$ in magnitude.

In the NNLO analysis the convergence of the series  
for the resonance energy can be essentially improved by 
employing  an infrared safe mass parameter instead 
of the top quark pole mass \cite{MelYel,BSS,Nag,HoaTeu2}.
As a consequence, using the value of  $\alpha_s$ with $2\%$ uncertainty  
as an input  the $\overline{MS}$ top quark mass
can be obtained from the resonance peak with the accuracy $\sim 100$ MeV. 
On the contrary, the behavior of the perturbative series for
the cross section normalization cannot be improved
in this way that makes the high precision determination 
of $\alpha_s$ and $\Gamma_t$ to be problematic.
At the same time the perturbation theory   behavior of the total cross 
sections of  $t\bar t$  production in  $e^+e^-$ annihilation 
and  $\gamma\gamma$ collision is very similar.
Thus the ratio of the cross sections taken, for example, at the resonance
energies  is expected to have very nice perturbative 
properties such as fast convergence  and stability against changing 
the  normalization scale. This quantity can probably be 
used to extract the value of $\alpha_s$
with rather high accuracy. To exploit this property
for the precision determination of the strong coupling constant 
one has to complete the calculation of the 
NNLO correction to the coefficient $C^{++}$
that now is the main challenge for the theoretical
study of the $\gamma\gamma\to t\bar t$  threshold production.

{\bf Acknowledgments.}
The author thanks the organizers of HEPC-2000 International
Workshop for making this fruitful meeting.
This work was supported in part by 
the Bundesministerium f\"ur Bildung und Forschung
under Contract No.\ 05~HT9GUA~3
and  by the Volkswagen Foundation under
Contract No.\ I/73611.

\newpage

\begin{center}
\setlength{\unitlength}{0.240900pt}
\ifx\plotpoint\undefined\newsavebox{\plotpoint}\fi
\sbox{\plotpoint}{\rule[-0.200pt]{0.400pt}{0.400pt}}%

\end{center}

\noindent
{\bf Fig. 1.}
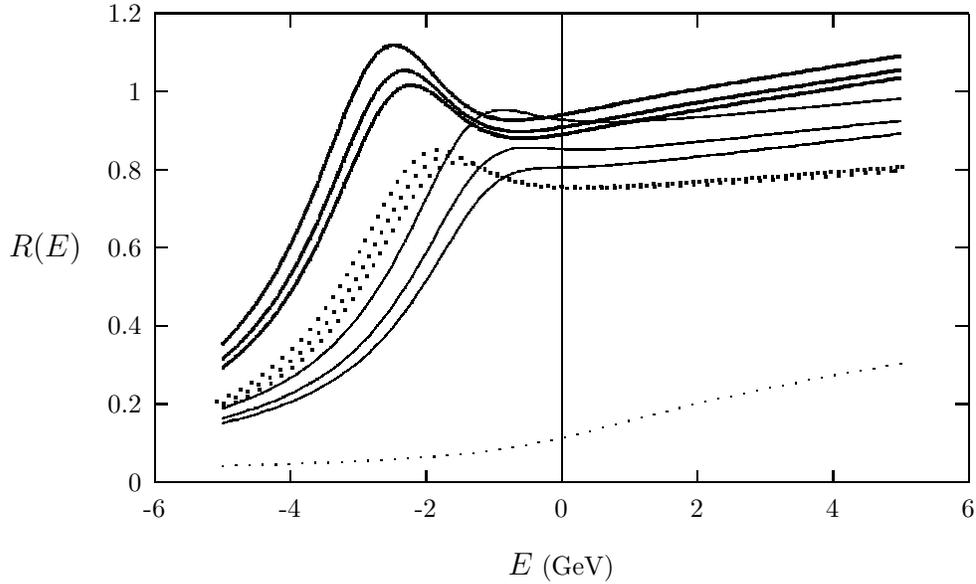
The normalized  cross section  $R(E)$
in the leading order (solid lines),
NLO (bold dotted lines) and
NNLO (bold solid lines) for $m_t=175~{\rm GeV}$, $\Gamma_t=1.43~{\rm GeV}$,
$\alpha_s(M_Z)=0.118$ and $\mu=50~{\rm GeV},~75~{\rm GeV}$
and $100~{\rm GeV}$. 
The dotted line corresponds to the result in Born approximation.

\vspace{10mm}

\begin{center}
\setlength{\unitlength}{0.240900pt}
\ifx\plotpoint\undefined\newsavebox{\plotpoint}\fi
\sbox{\plotpoint}{\rule[-0.200pt]{0.400pt}{0.400pt}}%
%
\end{center}

\noindent
{\bf Fig. 2.}
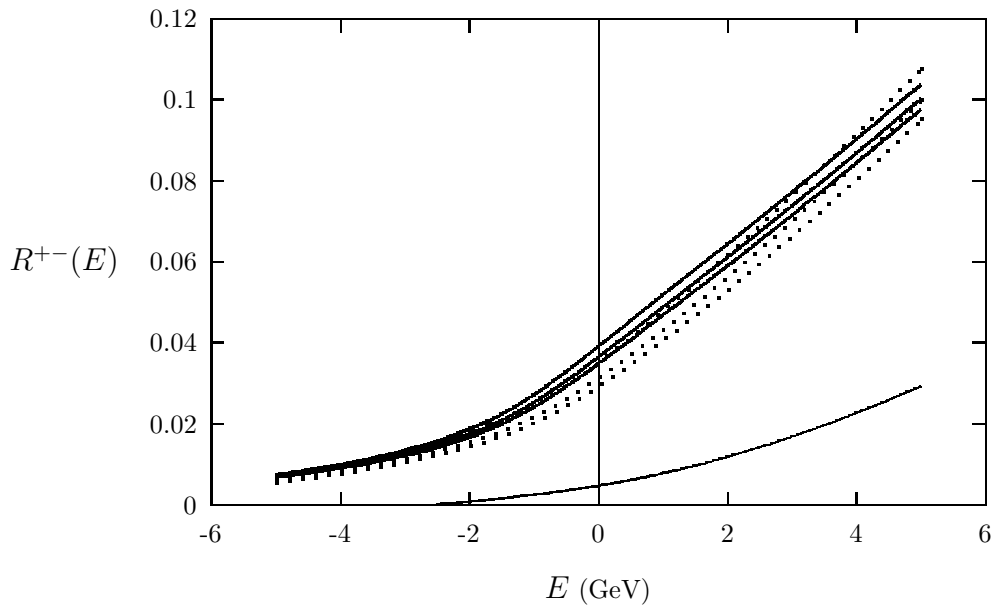
The normalized cross section  $R^{+-}(E)$
in the leading order (dotted lines)  and
NLO (bold solid lines) for $m_t=175~{\rm GeV}$, $\Gamma_t=1.43~{\rm GeV}$,
$\alpha_s(M_Z)=0.118$ and $\mu=50~{\rm GeV},~75~{\rm GeV}$
and $100~{\rm GeV}$.
The solid line corresponds to the result in Born approximation.

\end{document}